# Network Independent RIS: A Practical and Experimental Perspective

Emre Arslan*[‡], Fatih Kilinc*[§], Ertugrul Basar*, Huseyin Arslan[†]
*CoreLab, Department of Electrical and Electronics Engineering, Koç University, Sariyer 34450, Istanbul, Turkey
[†]Department of Electrical and Electronics Engineering, Istanbul Medipol University, 34810, Istanbul, Turkey
[‡] Turkcell 6GEN. Lab, Turkcell Iletisim Hizmetleri Inc., 34854, Istanbul, Turkey
[§]Department of Research and Development, ULAK Communications Inc., 06510, Ankara, Turkey
e-mail: emre.arslan@turkcell.com.tr, fatih.kilinc@ulakhaberlesme.com.tr, ebasar@ku.edu.tr, huseyinarslan@medipol.edu.tr

*Abstract*—The march towards 6G is accelerating and future wireless network architectures require enhanced performance along with significant coverage especially, to combat impairments on account of the wireless channel. Reconfigurable intelligent surface (RIS) technology is a promising solution, that has recently been considered as a research topic in standards, to help manipulate the channel in favor of users' needs. Generally, in experimental RIS systems, the RIS is either connected to the transmitter (Tx) or receiver (Rx) through a physical backhaul link and the RIS is controlled by the network and requires significant computation at the RIS for codebook (CB) designs. In this paper, we propose a practical user-controlled RIS system that is isolated from the network to enhance communication performance and provide coverage to the user based on the users location and preference. Furthermore, a low-complexity algorithm is proposed to aid in CB selection for the user which is performed through the wireless cloud to enable a passive and energy efficient RIS. Extensive experimental test-bed measurements demonstrate the enhanced performance of the proposed system while both results match and validate each other.

## I. INTRODUCTION

DRIVEN by the strict requirements and various applications for future 6G and beyond wireless communication systems, researchers from academia as well as industry have shown significant interest to the fairly new and developing but exceptionally promising reconfigurable intelligent surface (RIS) technology. RISs are comprised of a large number of small, low-cost, passive meta-material elements that acquire the ability to manipulate impinging signals such as reflect, absorb, and refract, hence allowing some sort of control over the wireless channel. RIS technology is seen as an alternative to beamforming techniques that require a large number of antennas to direct signals, without the need for additional buffering or processing. RISs may introduce phase shifts that can be controlled by software to impinging signals in an intelligent manner to enhance coverage [1], improve signal quality [2], [3], boost capacity [4], [5], and much more.

RISs have been utilized in novel ways and integrated with existing technologies to incorporate the flexibility and advantages of multiple technologies. While equalization is conventionally performed at the transmitter (Tx) or receiver (Rx) in frequency selective channels, the RIS is exploited to virtually equalize the channel over-the-air in [6]. In [7], non-orthogonal multiple access (NOMA) in the uplink is enabled over-the-air using an hybrid RIS where the RIS is partitioned to active and passive sections. Furthermore, coverage extension is investigated in [8], [9] while [10]–[12] provides solutions for energy efficiency for RIS systems without the use of additional RF chains. Generally, RISs function in the most ideal manner for cases where a direct line-of-sight does not exist or is weak [13], [14]. Another issue that must be taken into consideration for RIS systems is the multiplicative path loss of the Tx-RIS-Rx link, hence, making significant gains difficult especially in outdoor scenarios [13], [15].

The issues mentioned above are more prevalent in practical and experimental studies with limitations. RIS literature is becoming saturated from a theoretical and simulative perspective however, it is lacking and in need of practical and experimental studies especially since it is a study item envisioned to be in use commercially for future communication systems. There are currently ongoing pre-standardization efforts on ETSI to push RIS technology into 6G standards [16], [17]. Some experimental studies include [18] where the authors investigate the received power performance using horn antennas for an L-shaped indoor scenario and aim to enhance performance for blind spots. In [19] similarly a multi-bit RIS is deployed for coverage enhancement in an L-shaped indoor scenario with non line-of-sight (NLOS) with horn antennas. Additionally, a 1-bit RISs with 1100 elements for self- adaptive reflection is tested for indoor/outdoor environments in [20].

The authors in [21] aim to enhance indoor coverage using a physical RIS and also propose a codebook (CB) design

This work was supported by The Scientific and Technological Research Council of Turkey (TUBITAK) through the 1515 Frontier Research and Development Laboratories Support Program under Project 5229901 - 6GEN. Lab: 6G and Artificial Intelligence Laboratory, and also Grant 120E401. In addition it has been funded by, ULAK Communications Inc. E. Arslan, F. Kilinc and E. Basar are with the Communications Research and Innovation Laboratory (CoreLab), Department of Electrical and Electronics Engineering, Koç University, Sariyer 34450, Istanbul, Turkey. H. Arslan is with Communications Signal Processing and Networking Center, Department of Engineering and Natural Sciences, Istanbul Medipol University, 34810, Beykoz, Istanbul. E. Arslan and F. Kilinc are also with TURKCELL and ULAK, respectively. e-mail: earslan18@ku.edu.tr, fatih.kilinc@ulakhaberlesme.com.tr, ebasar@ku.edu.tr, huseyinarslan@medipol.edu.tr

without feedback channels. A few major points in this work that deteriorates the practicality of the system is that both the Tx and Rx use horn antennas and that they only consider a discrete number of training points where if you fall in between those exact points you suffer a significant reduction in received signal power. Furthermore, the system needs to train the RIS at each point through extensive measurements based on the Rx signal power where the Rx and RIS are connected to each other via a wire. The CB selection applied is based on trial and error of each individual RIS elements or a group of RIS elements turning on and off and observing the received signal power; which is impractical when it comes to daily use of customers requiring time and power.

Based on the works mentioned previously, generally RIS systems are designed in impractical ways to acquire significant gains and show improvement in communication. The use of horn antennas at both ends, extensive offline training before use of the RIS, physical back-haul links connecting the RIS to the Tx or Rx, unrealistic channel state information (CSI) requirements, exhaustive CB algorithms, and the RIS being dependent and connected to the network during online and offline use are just a few important factors that reduce the practicality of an RIS system. In this manuscript, the authors introduce a unique user-based perspective and aim to significantly improve the practicality of the RIS and isolate it from the network when in use. The user at the receiving end is equipped with an omni-directional antenna hence reflections from walls and obstacles around it will also be considered, which is typically not the case in most studies. There is no CSI required and no physical link between the receiving user and RIS, instead, feedback is provided through Wi-Fi which is available in most indoor locations. The RIS is completely independent from the Tx at all times and also when serving the receiving user. The only time the RIS is wirelessly connected to the Rx through Wi-Fi is when the indoor user requests to quickly update or add new CBs in an efficient manner. At any other case, the user sends short commands to the RIS only to change the RIS CB configuration to the users preference via online application/remote control. Additionally, we take into consideration the time for CB generation when requested by the user and propose an efficient and enhanced CB selection algorithm to quickly serve and boost user performance based on received signal power.

The contributions of this study can be summarized and highlighted as follows:

- A unique perspective to an end-to-end user-controlled RIS scenario is considered.
- A practical user-controlled RIS system isolated from the network is proposed.
- A novel CB selection algorithm is proposed to favor a passive RIS with efficiency, low complexity, and enhanced performance.
- Extensive experimental results with a test-bed are provided to compliment and validate the effectiveness of the proposed CB algorithm and RIS scenario.

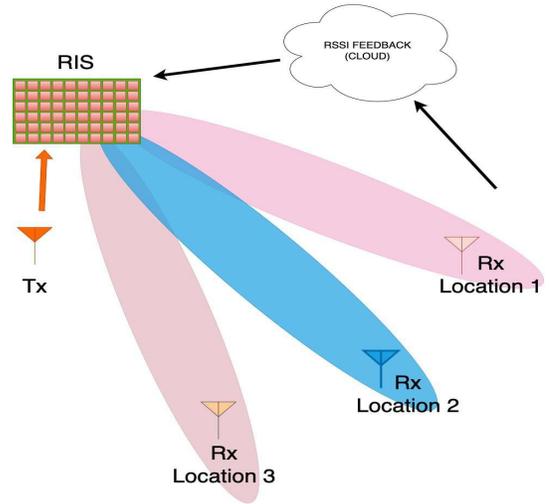

Fig. 1: Comparison of exact and simulated results for varying number of RIS elements $N$.

The rest of the paper can summarized as follows. In Section II, the system framework of the proposed RIS scheme is presented. In Section III the proposed CB selection algorithm is explained. Section IV presents the test-bed set up subsequently, the experimental analysis discussed in Section V. Finally, Section VI concludes the paper.

## II. SYSTEM FRAMEWORK

This section provides the technical details of the proposed RIS assisted wireless communication system setup. First, the system and signal model is presented followed by the setup of the experimental test-bed.

### A. System Model

In this study, communication occurs between a single-antenna Tx and Rx under the aid of a passive RIS with $N = C \times R$ electromagnetic elements, where $C$ and $R$ denotes the number of columns and rows of the RIS, respectively. Each individual element of the RIS has the ability to introduce $0°$ or $180°$ phase shifts in both horizontal and vertical polarization, corresponding to off and on states, to the incident signal from the Tx. Since an RIS is employed generally in weak LOS or NLOS systems, we also consider a NLOS system where the channel from the Tx to the RIS is defined as $\mathbf{h} \in \mathbb{C}^{N \times 1}$ and the channel from the RIS to the Rx is defined as $\mathbf{g} \in \mathbb{C}^{N \times 1}$. As illustrated in Fig. 1, the RIS serves the receiving user from multiple locations depending on the users preference and selection from his/her application or remote control. The $k^{\text{th}}$ complex baseband signal transmitted by the Tx is defined as $x[k]$ hence, and the received signal $r[k]$ can be expressed as follows:

$$r[k] = (\mathbf{g}^{\text{T}} \mathbf{\Phi}_{\text{RIS}} \mathbf{h}) x[k] + n[k] \qquad (1)$$

where $\mathbf{\Phi}_{RIS} = \text{diag}(\alpha_1 e^{j\phi_1}, \cdots, \alpha_N e^{j\phi_N})$ denotes the phase matrix at the RIS and $n[k]$ represents the additive white Gaussian noise (AWGN) component at the receiver caused by hardware components which follows the distribution $\mathcal{CN}(0, \sigma_n^2)$. In the RIS phase matrix or in other words the CB, $\alpha_n$ is the magnitude of the reflection coefficient of the $n^{th}$ RIS element and is constant for a passive RIS, whereas the phase introduced by the $n^{th}$ RIS element is $\phi_n$ and can take the values $0°$ and $180°$, for $n \in \{1, \cdots, N\}$.

RISs are typically used to significantly improve coverage and the signal-to-noise ratio at the Rx by either eliminating interference, coherently aligning phases, or other novel methods by manipulating the channel. By adjusting the RIS CB $\mathbf{\Phi}_{RIS}$ we can increase the end-to-end gain $|\mathbf{g}^T \mathbf{\Phi}_{RIS} \mathbf{h}|^2$ thus proportionally increasing the SNR of the received signal. The optimal CB that maximizes the gain is expressed as follows:

$$\mathbf{\Phi}_{RIS}^* = \arg\max_{\phi_n, \forall n} \quad |\mathbf{g}^T \mathbf{\Phi}_{RIS} \mathbf{h}|^2$$
$$\text{s.t.} \quad \phi_n \in \{0, 180\}, \forall n. \quad (2)$$

To solve this optimization problem, other studies conduct statistical channel estimation methods however, we avoid complex techniques to obtain CSI and aim for low complexity for practicality. Another approach is to perform a heuristic search for the optimum CB which maximizing the gain. Since each individual element can be configured on and off in both polarizations resulting in four possible states, the search space is $4^{76}$. Since $4^{76}$ is a large search space and searching it iteratively would increase the complexity of the system, we need an algorithm do search this space as fast and efficient as possible without degrading performance significantly. Hence, in the next section, we will propose a novel CB algorithm to search this space efficiently and even enhance the performance.

In the proposed system, we focus on maximizing the received power at the end user based on the maximization problem as provided in (2). Optimization of the CB that directs the incident signal from the Tx to the Rx and provides maximum received signal power in a fast, efficient and practical way is prioritized.

### III. PROPOSED CODEBOOK ALGORITHM

In this section, the proposed CB algorithm is introduced and discussed in detail. Generally, CB search algorithms or methods in existing RIS literature depend on offline training, iterative and complex solutions. The focus of the proposed CB design is to provide a low-complex, fast and practical CB construction based on RSSI feedback from the user-end device. In [21], the authors turn the RIS elements on and off one by one and decide to keep the RIS element on or off based on RSSI feedback. Although this method provides decent performance, it needs to be done iteratively at each location in an offline manner and if a user falls in between these trained points, performance will degrade significantly when in online use. On the other hand, the authors from [22] propose a low complexity algorithm doing a horizontal search first then on that trained CB, perform a vertical search on top of it. Though this is reduced in complexity, it is significantly degraded in performance and can be further enhanced without a drastic trade-off in the system performance. To introduce a more practical and efficient approach we consider the algorithm design presented in Algorithm 1.

**Algorithm 1 : Influential Element based Iterative Algorithm**

1: **Input:** $C, R$, RSSI
2: $\mathbf{\Phi}_{RIS}, \mathbf{\Phi}_h, \mathbf{\Phi}_v, \mathbf{\Phi}_{temp} \leftarrow [0, \cdots, 0]$ // Set all RIS elements to OFF state
3: $\text{RSSI} \leftarrow P_r^0$  // Measuring initial received power
4: $P_r^{\max} \leftarrow P_r^0$  // Setting maximum power to initial power
5: // Horizontal Search
6: **for** $r = 1 : R$ **do**  // Iterative Row Search
7:    **for** $s = 1 : S$ **do**
8:       $\mathbf{\Phi}_{temp}[\phi_{r,1}, \cdots, \phi_{r,C}] \leftarrow S[s]$
9:       $P_r^c \leftarrow \text{RSSI}(\mathbf{\Phi}_{temp})$  // RSSI feedback
10:      **if** $P_r^c > P_r^{\max}$ **then**
11:         $\mathbf{\Phi}_h \leftarrow \mathbf{\Phi}_{temp}$
12:         $P_r^{\max} \leftarrow P_r^c$
13:      **end if**
14:    **end for**
15:    $\mathbf{\Phi}_{temp} \leftarrow \mathbf{\Phi}_h$
16: **end for**
17: **return** $\mathbf{\Phi}_h$
18: $P_r^{\max} \leftarrow P_r^0$  // Setting maximum power back to initial power
19: $\mathbf{\Phi}_{temp} \leftarrow [0, \cdots, 0]$  // Setting all RIS elements back to OFF state
20: // Vertical Search
21: **for** $c = 1 : C$ **do**  // Iterative Column Search
22:    **for** $s = 1 : S$ **do**
23:       $\mathbf{\Phi}_{temp}[\phi_{1,c}, \cdots, \phi_{R,c}] \leftarrow S[s]$
24:       $P_r^c \leftarrow \text{RSSI}(\mathbf{\Phi}_{temp})$  // RSSI feedback
25:      **if** $P_r^c > P_r^{\max}$ **then**
26:         $\mathbf{\Phi}_v \leftarrow \mathbf{\Phi}_{temp}$
27:         $P_r^{\max} \leftarrow P_r^c$
28:      **end if**
29:    **end for**
30:    $\mathbf{\Phi}_{temp} \leftarrow \mathbf{\Phi}_v$
31: **end for**
32: **return** $\mathbf{\Phi}_v$
33: $\mathbf{\Phi}_{RIS} \leftarrow \mathbf{\Phi}_h \cap \mathbf{\Phi}_v$  // Set influential elements
34: **procedure** ITERATION OVER REMAINING ELEMENTS
35:    $P_r^{\max} \leftarrow \text{RSSI}(\mathbf{\Phi}_{RIS})$  // RSSI feedback
36:    $\mathbf{\Phi}_{temp} \leftarrow \mathbf{\Phi}_{RIS}$
37:    **for** $j \in \mathbf{\Phi}_{RIS} \land j \in / (\mathbf{\Phi}_h \cap \mathbf{\Phi}_v)$ **do**
38:       **for** $s = 1 : S$ **do**
39:         $\mathbf{\Phi}_{temp}[j] \leftarrow S[s]$
40:         $P_r^c \leftarrow \text{RSSI}(\mathbf{\Phi}_{temp})$  // RSSI feedback
41:         **if** $P_r^c > P_r^{\max}$ **then**
42:           $\mathbf{\Phi}_{RIS} \leftarrow \mathbf{\Phi}_{temp}$
43:           $P_r^{\max} \leftarrow P_r^c$
44:         **end if**
45:       **end for**
46:       $\mathbf{\Phi}_{temp} \leftarrow \mathbf{\Phi}_{RIS}$
47:    **end for**
48:    **return** $\mathbf{\Phi}_{RIS}$
49: **end procedure**

In Algorithm 1 we first initialize CB instances $\mathbf{\Phi}_{RIS}, \mathbf{\Phi}_h, \mathbf{\Phi}_v$, and $\mathbf{\Phi}_{temp}$ by setting all elements to the off state and the configurations denote the CBs for overall, horizontal, vertical and temporary configurations, respectively. The variables $P_r^0, P_r^{\max}$, and $P_r^c$ correspond to the initial, maximum and current RSSI values. First, we take RIS elements row by row and based on RSSI feedback we decide whether the elements of that row should be on or off in between the Lines 6-17. The first for loop iterates over the elements in the same row, by applying $S = 4$ different states in Lines 6 and 7. $\mathbf{\Phi}_{temp}$ is updated such the $s$th state is applied to all elements on the $r$th row in Line 9. $P_r^c$ is updated based on the temporary CB, then if $P_r^c$ exceeds $P_r^{\max}$, it means that the applied states contribute to the overall system gain. Therefore, we update $\mathbf{\Phi}_h$ if a temporary CB is beneficial and update $P_r^{\max}$ accordingly in Lines 26-27. Finally, the best horizontal configuration is obtained and returned in Line 31. After resetting $P_r^{\max}$ to $P_r^0$ and $\mathbf{\Phi}_{\text{temp}}$ to fully off state, the vertical search is conducted by following the similar steps in between the Lines 21-32.

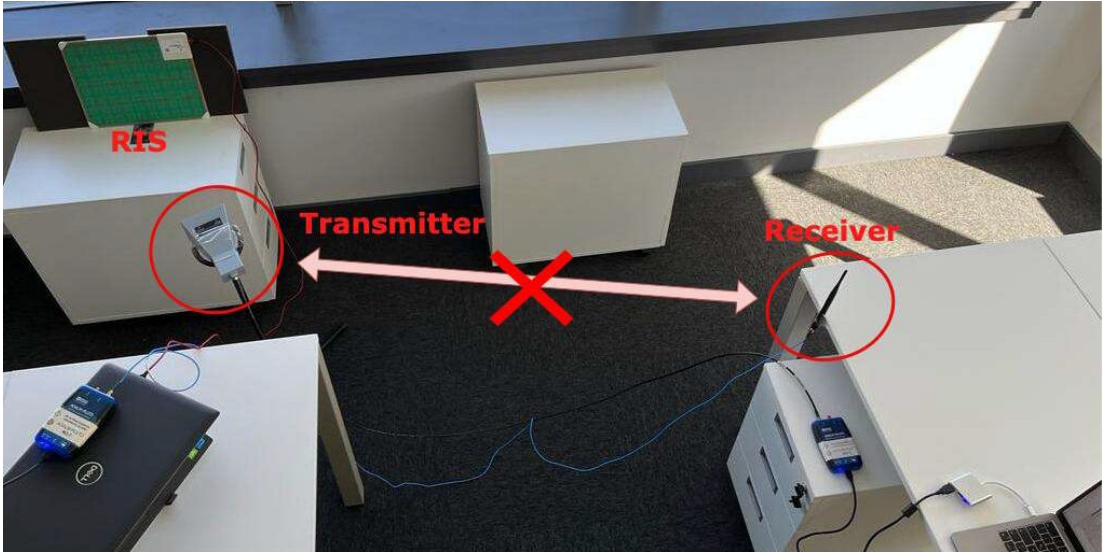

Fig. 2: Test-bed set-up for the experimental analysis.

Then we look at both horizontal and vertical cases and select the elements that have the same state and identify the most influential elements i.e. on/off in both horizontal and vertical cases. The most influential elements are decided by finding the intersecting elements with same states between $\mathbf{\Phi}_h$ and $\mathbf{\Phi}_v$ in Line 33. It should be noted that a significant amount of RIS elements overlap and are identified as influential elements and with the increase of these elements, the complexity decreases. In addition, as the RIS size increases, or polarization decreases the number of influential elements increase, hence, reducing complexity even further. After deciding the most influential elements and setting their states for $\mathbf{\Phi}_{\text{RIS}}$, we start to build the states of the remaining elements in the procedure defined as in [21] in Line 34. Since we are building the CB on top of the influential elements, $P_r^{\max}$ is set to RSSI value based on $\mathbf{\Phi}_{\text{RIS}}$. The loops in Line 37 and 38 iterates over the remaining elements one by one, by applying $S = 4$ different states. The search for the states of remaining elements is conducted by following the similar steps as explain in horizontal search, however this time the iteration goes over the remaining elements one by one in Line 37, by applying $S = 4$ different states in Line 38 rather than horizontal group of elements. Consequently, the final CB $\mathbf{\Phi}_{\text{RIS}}$ which maximizes the RSSI is returned.

While the bench marks from [21] and [22] have a complexity of $O(4 \times C \times R)$ and $O(C + R)$, the proposed scheme has a complexity of $O(C + R + 4 \times (C \times R - I))$ where $I = |\mathbf{\Phi}_h \cap \mathbf{\Phi}_v|$ is equal to number of influential elements. We propose a reduced complexity from the iterative method in [21] while providing superior performance unlike [22]. It should be noted that the proposed algorithm is less complex in cases where $I >> C + R$, which occurs most of the time. Furthermore, when the RIS size increases and a single polarization is considered, the likelihood of a lower complexity is guaranteed.

When the receiving user wants to set up a new CB, they can go to the desired location and quickly run this algorithm and save it on the RIS while online. Once a CB is saved, the user can just select that saved CB whenever they are at that location. There is no need for extensive offline training. Furthermore, the proposed CB takes into consideration of the impact of neighboring elements since we use a horizontal and vertical approach. Since we find the most significant elements and then do an iterative approach to the remainder of elements, the proposed CB algorithm not only provides low-complexity but also provides enhanced signal power hence, improved communication performance.

## IV. TEST-BED SET-UP

In this section we will discuss the set up of our system and details of the equipment used. Furthermore, the control mechanisim and end-to-end user experience process will be discussed.

Fig. 2 presents the general model of our set up for the experimental analysis. As seen in Fig. 2 the TX is a horn antenna with a half beamwidth of $40°$, on the other hand for a practical approach, the receiver is equipped with an omni-directional antenna. While most literature consider specific geometric set-up cases such as L-shaped hallways [21], we consider an indoor office environment where the RIS is mounted on a wall or the edge of the room and the end-user equipped with an omni-directional antenna is located at different desks which are positioned at various ends in the office considering numerous geometric and reflective scenarios. Through the aid of two ADALM-PLUTO software-defined radio (SDR) modules, the signal $x[k]$ is modulated with $M$-PSK and transmission occurs at 5.2 GHz with −10 dBm power with a horn antenna directed towards the RIS to ensure no line-of-sight between the Tx and Rx, on the other hand, reception takes place in different locations of the office with an omni-directional antenna measuring the received signal power at 5.2 GHz. A 10 × 8 uniform planar array RIS prototype designed by Greenerwave [23] is exploited to redirect impinging signals towards the desired area via software-controlled phase shifts. Operating at a center frequency of 5.2 GHz and working bandwidth of 1 GHz, the RIS has 76 controllable elements via pin diodes, that have the ability to introduce 0 or 180 phase shifts in both the horizontal and vertical polarizations. Hence, the RIS consists of $4^{76}$ possible configurations resulting in a significantly inefficient exhaustive search algorithm [24].

In our system, the RIS is independent from the Tx and Rx and initially has all elements set to 0 deg. When the receiving user goes to a desired location and wants to run the proposed algorithm through the use of an app or remote control, the user selects new configuration and requests to run the CB algorithm. When the CB algorithm is initiated, the RIS and receiver connects to the cloud/server. As the states of the RIS elements change and search for a new configuration, the server requests RSSI feedback from the receiver. Based on the request, the receiving device captures 50 frames and with $10^6$ samples per frame and feedbacks the average RSSI value. Hence, the proposed algorithm quickly produces a sub-optimal CB for that location through the cloud and saves it at the RIS. The RIS then disconnects from the network and passively serves the user. This can be done for numerous locations which will be denoted as Loc A, Loc B, Loc C, and the CB for each location will be stored at the RIS. The user can delete or add new CBs as desired and through the use of the remote select whichever CB to apply at the RIS. Hence, the RIS is passive and isolated from the network and only connects to the cloud momentarily to run the practical and high performance CB generation algorithm when desired.

## V. EXPERIMENTAL ANALYSIS

In this section we present the experimental results and compare the performance of the proposed system with both benchmark 1 from [21] and benchmark 2 from [22] as mentioned in Section 3.

Fig. 3 displays the performances and complexity of the proposed system and benchmarks through the received signal power and number of iterations, respectively. As it can be deducted, the proposed CB algorithm presents a desirable outcome providing superior performance than both benchmarks. When compared to benchmark 1, the proposed algorithm exceeds in both complexity and performance. On the other hand, when compared to benchmark 2, the proposed system has

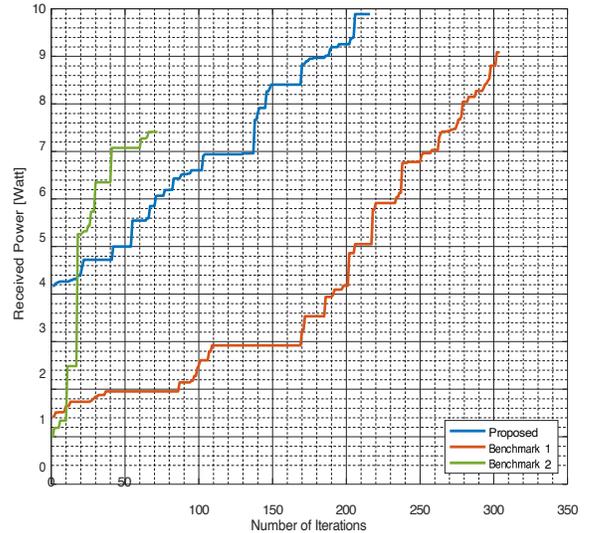

Fig. 3: Performance comparison between the proposed CB algorithm and benchmarks.

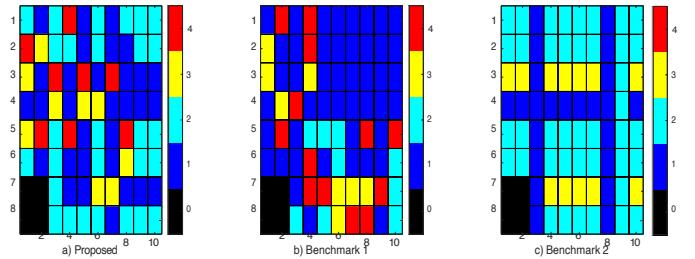

Fig. 4: CBs generated from the proposed algorithm and benchmark algorithms.

higher complexity but is admirable in its superior performance. It should be noted that the proposed scheme

Fig. 4 shows the CBs generated on the cloud and applied at the RIS. The black color indicates the controller part of the RIS which doesn't consist of reflecting RIS elements. In the configuration instances for each of the RIS elements, the dark and light blue, and the yellow and red colors represent the 0° and 180° states in the horizontal and vertical polarization, respectively.

In Fig. 5, we place the receiving user at three varying locations and generate a CB for each. The three locations are denoted by Loc A, Loc B and Loc C, and the configured and stored CBs for these locations are denoted by CB-A, CB-B and CB-C, respectively. Then the user moves to each location and all generated CBs are applied in succession and the performance in received signal power is measured. It can be observed that the CB generated for the corresponding location has a significant impact on the user performance while the other CBs generated at other locations in the office do not provide similar gains. This behaviour shows that the proposed CB algorithm works accurately and stored CBs perform based on the user selected location.

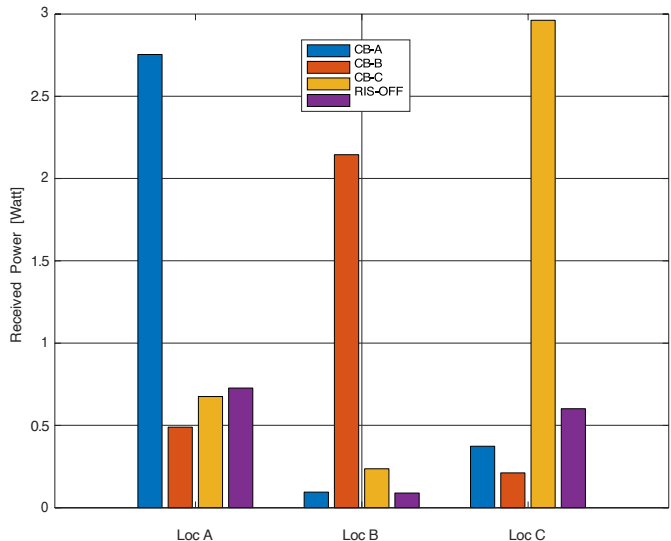

Fig. 5: CB performances at varying locations.

## VI. Conclusion

In this study, a unique and practical user controlled RIS system and perspective is considered. In addition, the authors propose a novel, practical and efficient CB algorithm while reducing the complexity and also enhancing the performance of the RIS aided system. Extensive experimental results are demonstrated, discussed and confirm the performance enhancement along with the practicality of the system design. Future directions may include applications of machine learning the further improve performance and complexity of the proposed system design.